\documentclass[11pt]{article}
\usepackage[letterpaper, margin=1in]{geometry}
\usepackage[utf8]{inputenc}
\def\PrintMode{0}%
\newif\ifmynotes
\mynotestrue

\usepackage{tikz}
\usetikzlibrary{positioning}
\usepackage{pgffor}
\usetikzlibrary{decorations.pathreplacing}
\usepackage{expl3}
\usepackage[T1]{fontenc}
\usepackage{graphicx}
\usepackage{float}
\usepackage[lined,ruled,linesnumbered]{algorithm2e}
\usepackage{xspace}
\usepackage[utf8]{inputenc}
\usepackage[noadjust]{cite}
\usepackage{amsmath, amssymb, amsfonts, amsthm}
\usepackage{enumerate}
\usepackage{xcolor}
\usepackage[all]{xy}
\usepackage[linktocpage=true,pagebackref=true,colorlinks,linkcolor=magenta,citecolor=blue,bookmarks,bookmarksopen,bookmarksnumbered]{hyperref}
\usepackage{bm}
\usepackage{bbm}
\usepackage[normalem]{ulem}
\usepackage{paralist}
\usepackage{gensymb}
\usepackage{thmtools}
\usepackage{rotating}
\usepackage{mdframed}
\usepackage{multirow}
\usepackage{appendix}
\usepackage{tabularx}
\usepackage{verbatim}
\usepackage{mathrsfs}
\usepackage{indentfirst}
\usepackage{mleftright}
\usepackage[nottoc]{tocbibind}%
\usepackage{complexity}
\usepackage{tablefootnote}
\usepackage{epigraph}
\usepackage{autobreak}

\usepackage{todonotes}

\ifnum\PrintMode=1

\usepackage{savetrees}
\else
\fi

\graphicspath{{./figs/}}

\usepackage[capitalize]{cleveref}%
\newtheorem{theorem}{Theorem}[section]
\newtheorem{lemma}[theorem]{Lemma}

\newtheorem{corollary}[theorem]{Corollary}
\newtheorem{claim}[theorem]{Claim}

\newtheorem{conjecture}[theorem]{Conjecture}

\theoremstyle{definition}
\newtheorem{definition}[theorem]{Definition}

\newtheorem{remark}[theorem]{Remark}
\newtheorem*{remark*}{Remark}

\renewcommand*\backref[1]{\ifx#1\relax \else (cit.~on p.~#1) \fi} %

\usetikzlibrary{patterns}

\makeatletter
\def\moverlay{\mathpalette\mov@rlay}
\def\mov@rlay#1#2{\leavevmode\vtop{%
		\baselineskip\z@skip \lineskiplimit-\maxdimen
		\ialign{\hfil$\m@th#1##$\hfil\cr#2\crcr}}}
\newcommand{\charfusion}[3][\mathord]{
	#1{\ifx#1\mathop\vphantom{#2}\fi
		\mathpalette\mov@rlay{#2\cr#3}
	}
	\ifx#1\mathop\expandafter\displaylimits\fi}
\makeatother

\renewcommand{\poly}{\mathrm{poly}}

\newlang{\MCSP}{MCSP}
\newlang{\MFSP}{MFSP}
\newlang{\MKtP}{MKtP}
\newlang{\MKTP}{MKTP}
\newlang{\itrMCSP}{itrMCSP}
\newlang{\itrMKTP}{itrMKTP}
\newlang{\itrMINKT}{itrMINKT}
\newlang{\MINKT}{MINKT}
\newlang{\MINK}{MINK}
\newlang{\MINcKT}{MINcKT}
\newlang{\CMD}{CMD}
\newlang{\DCMD}{DCMD}
\newlang{\CGL}{CGL}
\newlang{\PARITY}{PARITY}
\newlang{\Empty}{\textsc{Empty}}
\newlang{\Avoid}{\textsc{Avoid}}
\newlang{\Sparsification}{\textsc{Sparsification}}
\newlang{\HamEst}{\mathsf{HammingEst}}
\newlang{\HamHit}{\mathsf{HammingHit}}
\newlang{\CktEval}{\textsc{Circuit-Eval}}
\newlang{\Hard}{\textsc{Hard}}
\newlang{\cHard}{\textsc{cHard}}
\newlang{\CAPP}{CAPP}
\newlang{\GapUNSAT}{GapUNSAT}
\newlang{\OV}{OV}
\renewlang{\PCP}{PCP}
\newlang{\PCPP}{PCPP}
\newclass{\Avg}{Avg}
\newclass{\ZPEXP}{ZPEXP}
\newclass{\DLOGTIME}{DLOGTIME}
\newclass{\ALOGTIME}{ALOGTIME}
\newclass{\ATIME}{ATIME}%
\newclass{\SZKA}{SZKA}
\newclass{\Laconic}{Laconic\text{-}}
\newclass{\APEPP}{APEPP}
\newclass{\SAPEPP}{SAPEPP}
\newclass{\TFSigma}{TF\Sigma}
\newclass{\NTIMEGUESS}{NTIMEGUESS}

\newlang{\Formula}{Formula}
\newlang{\THR}{THR}
\newlang{\EMAJ}{EMAJ}
\newlang{\MAJ}{MAJ}
\newlang{\SYM}{SYM}
\newlang{\DOR}{DOR}
\newlang{\ETHR}{ETHR}
\newlang{\Midbit}{Midbit}
\newlang{\LCS}{LCS}
\newlang{\TAUT}{TAUT}
\newlang{\Poly}{\text{-}Poly}

\newcommand{\F}{\mathbb{F}}

\DeclareMathOperator*{\Span}{\mathrm{Span}}

\renewcommand{\epsilon}{\varepsilon}
\newcommand{\eps}{\epsilon}

\usepackage[draft, inline,marginclue,index]{fixme}  %
\fxsetup{theme=color,mode=multiuser}

\definecolor{color1}{RGB}{46,134,193}

\definecolor{color7}{RGB}{128,0,128}
\FXRegisterAuthor{yeyuan}{ayeyuan}{\color{red}Yeyuan}

\definecolor{color3}{RGB}{255,128,0}
\FXRegisterAuthor{zihan}{azihan}{\color{blue}Zihan}

\FXRegisterAuthor{kingsiong}{akingsiong}{\color{violet}King-Siong}

\definecolor{color5}{RGB}{128,128,128}
\FXRegisterAuthor{task}{atask}{\colorbox{color5}{\color{white}Task}}

\newif\ifmynotes
\mynotestrue

\title{Evidence Toward a Conjecture in List Recoverability of Folded Reed--Solomon Codes}

\author{Yeyuan~Chen\thanks{University of Michigan, Ann Arbor. E-mail: \texttt{yychen9961@gmail.com}.}
\and Mahdi~Cheraghchi\thanks{University of Michigan, Ann Arbor. E-mail: \texttt{mahdich@umich.edu}.}
\and King-Siong~Si\thanks{National University of Singapore. E-mail: \texttt{kingsiong.tcs@gmail.com}.}
\and Zihan~Zhang\thanks{Simons Institute for the Theory of Computing, University of California, Berkeley. E-mail: \texttt{zihanz@hku.hk}.}}

\date{}
\begin{document}
\maketitle
\begin{abstract}
This paper investigates the list-recoverability of folded Reed--Solomon codes and provides new evidence toward a recent conjecture. Chen--Zhang (STOC 2025) established that for any $\ell,a \ge 2$ and $L = \ell^a - 1$, if a folded Reed--Solomon code of rate $R$ is $(\rho,\ell,L)$-list-recoverable, then $\rho\leq \frac{L+1-\ell}{L+1}(1-\frac{aR}{a-1}).$
It is conjectured that this upper bound is tight. We provide two pieces of evidence concerning this conjecture. First, we prove that explicit folded Reed--Solomon codes of rate arbitrarily close to $\tfrac{a-1}{a+\ell-2}$ are $(0,\ell,L)$-list-recoverable, which confirms the conjecture in the special case $\rho=0$ and $\ell=2$. Second, we show that for any $R\in(0,\tfrac{a-1}{a})$, there are rates arbitrarily close to $R$ at which attaining the conjectured bound requires exponentially large alphabets. This mirrors the phenomenon observed by Alrabiah--Guruswami--Li (SODA 2024), where achieving the generalized Singleton bound in the list-decoding setting also necessitates exponential alphabet size.
\end{abstract}

\section{Introduction}
Error-correcting codes facilitate reliable communication over noisy channels by mapping messages $m \in \Sigma^k$ to codewords $c \in \Sigma^n$. A code $\mathcal{C} \subseteq \Sigma^n$ with rate $R = k/n$ and minimum (relative) Hamming distance $\delta$ ensures correct decoding over an adversarial channel introducing up to a $\delta/2$-fraction of symbol errors. To tolerate a larger fraction of corruptions, one relaxes unique decoding to \emph{list decoding}, wherein the decoder outputs all codewords within a prescribed Hamming radius around the received word (e.g., \cite{sudan1997decoding,guruswami1998improved,shangguan2020combinatorial,brakensiek2023generic,gz23,agl24}).

A natural extension of list decoding is \emph{list recovery}, which is not only more difficult but also finds broader use in areas such as pseudorandomness \cite{ta2007lossless,guruswami2009Unbalanced,ta2012better,kalev2022unbalanced} and algorithms \cite{indyk2010efficiently,ngo2011efficiently,larsen2019heavy,dw22}. Formally, for any code $\mathcal{C}\subseteq \Sigma^n$, we say $\mathcal{C}$ is $(\rho, \ell, L)$-list-recoverable if, for any product set $S=S_1\times\cdots\times S_n$ with $S_i\subseteq\Sigma$ and $|S_i|\leq\ell$ for every $i\in[n]$, there are at most $L$ codewords $c \in \mathcal{C}$ such that
\[\mathrm{dist}(c,S)\leq\rho n,\]
where $\mathrm{dist}(c,S)$ denotes the number of indices $i\in[n]$ such that $c_i\notin S_i$.

Much like in the list decoding setting, a Johnson-type bound for list recovery \cite{gs01} guarantees that any maximum distance separable code is $(1 - \sqrt{\ell R}, \ell, \poly(n))$-list-recoverable. Furthermore, the notion of list-recovery capacity has been established in \cite{gi01, guthesis, resthesis}.
Let $q \geq \ell \geq 1$, $0 < \rho < 1 - \frac{\ell}{q}$, and $\epsilon > 0$. 
As stated in \cite[Theorem 2.4.12]{resthesis}, for sufficiently large $n$, random codes of block length $n$, rate $R$, and alphabet size $q$ are $(\rho, \ell, O\left(\frac{\ell}{\epsilon}\right))$-list-recoverable with high probability, where \begin{equation}\label{recovercapacity} R \geq 1 - H_{q, \ell}(\rho) - \epsilon, \end{equation} and $H_{q, \ell}(\rho) := \rho \log_q{\left(\frac{q - \ell}{\rho}\right)} + (1 - \rho) \log_q{\left(\frac{\ell}{1 - \rho}\right)}$ is the $(q, \ell)$-ary entropy function defined in \cite[Definition 2.4.9]{resthesis}. Codes satisfying \eqref{recovercapacity} are said to achieve list-recovery capacity. When $q \geq 2^{\Omega\left(\log{\ell}/\epsilon\right)}$, condition \eqref{recovercapacity} simplifies to $\rho \geq 1 - R - \epsilon$.

In this paper, we focus on the list-recoverability of folded Reed--Solomon (RS) codes. 
Folded RS codes, independently introduced by Krachkovsky and Guruswami--Rudra in \cite{krachkovsky2003reed,guruswami2008explicit}, are a simple variant of Reed--Solomon codes \cite{reed1960polynomial}. Formally, given any parameters $s,n,k>0$, a finite field $\mathbb{F}_q$, where $|\mathbb{F}_q|>sn>k$, a generator $\gamma$ of $\mathbb{F}^\times_q$, and any sequence of $n$ elements $\alpha_1,\alpha_2,\dots,\alpha_n\in\F_q$, the corresponding $(s,\gamma)$-folded RS code over the alphabet $\mathbb{F}_q^s$ with block length $n$ and rate $R=\frac{k}{sn}$ is defined as 
\[\mathsf{FRS}^{(s,\gamma)}_{n,k}(\alpha_1,\alpha_2,\dots,\alpha_n):=\bigg\{\mathcal{C}(f):f\in\mathbb{F}_q[X],\text{ }\deg f<k\bigg\}\subseteq\left(\mathbb{F}_q^s\right)^n,
\]
where $\mathcal{C}(f):=(F_1,F_2,\dots,F_n)\in\left(\mathbb{F}_q^s\right)^n$, with $F_i:=\bigl(f(\alpha_i),f(\gamma\alpha_i),\ldots,f(\gamma^{s-1}\alpha_i)\bigr)$, is the encoding of $f$. Note that RS codes correspond to the special case of $s=1$. An $(s,\gamma)$-folded RS code has \emph{appropriate} evaluation points if the $sn$ elements $\gamma^i\alpha_j$ are distinct, where $i\in\{0,\dots,s-1\}$ and $j\in[n]$.

A long line of work has made exciting progress on the list-recoverability of randomly punctured and explicit RS codes, as well as folded RS codes, beyond the Johnson radius \cite{Guru06,guruswami2013linear,kopparty2018improved,kopparty2023improved,lp20,guo2022improved,goldberg2022list,tamo2024tighter,CZ25,levi2024random,li2025near}.

On the positive side, for explicit folded RS codes, \cite{Guru06} first proved their $(1-R-\epsilon,\ell,\poly(n))$-list-recoverability. This was later refined in \cite{guruswami2013linear}, which showed that explicit $(s,\gamma)$-folded RS codes are $(\frac{1}{L+1}(L+1-\ell-\frac{sLR}{s-L+1}), \ell, \poly(n))$-list-recoverable with the candidate messages confined to an affine subspace of dimension at most $L-1$. Building on this framework, subsequent works \cite{kopparty2018improved,kopparty2023improved,tamo2024tighter} reduced the list size to a constant. In particular, they showed that explicit folded RS codes are $(1-R-\epsilon,\ell,(\tfrac{\ell}{\epsilon})^{O((1+\log \ell)/\epsilon)})$-list-recoverable.

On the negative side, Goldberg, Shangguan, and Tamo \cite{gst22b} first showed that for any $\ell \geq 2$ and $\epsilon > 0$, any code with sufficiently large block length $n$ and constant rate $R$ cannot be $(1 - R - \epsilon, \ell, \frac{\ell(1 - R)}{\epsilon} - 2)$-list-recoverable. However, while random codes achieve list-recovery capacity, a recent result of Chen and Zhang \cite{CZ25} proves that RS and folded RS codes cannot be $(1-R-\epsilon,\ell,\ell^{o_R(1/\epsilon)})$-list-recoverable. This rules out the possibility that such codes could achieve the list-recovery capacity $(1-R-\epsilon,\ell,O(\ell/\epsilon))$.
Based on their ideas, similar separation results on list recovery were extended to random linear codes \cite{levi2024random} and arbitrary linear codes \cite{li2025near}, respectively. 

Although folded RS codes cannot achieve list-recovery capacity, there remains a substantial gap between the best known upper and lower bounds. Motivated by potential applications such as the construction of good condensers \cite{guruswami2009Unbalanced}, it remains highly valuable to determine the optimal trade-off between the radius $\rho$, input size $\ell$, list size $L$, and rate $R$ for list recovery of folded RS codes. From the negative side, Chen and Zhang \cite{CZ25} proved that for any $\ell,a\ge 2,L=\ell^a-1$, achieving list-recoverability beyond $(\frac{L+1-\ell}{L+1}(1-\frac{aR}{a-1}),\ell,L)$ is asymptotically impossible. It is conjectured in \cite{CZ25} that this bound is actually tight, which means it is possible to achieve list-recoverability arbitrarily close to this bound. The conjecture is formulated as follows.

\begin{conjecture}[See Conjecture 3.6 \cite{CZ25}]\label{conj:main}
    For any constants $\epsilon>0, \ell\ge 2, L+1=\ell^a, a\in\mathbb{N}^{\ge 2}, R\leq \frac{a-1}{a}$ and a generator $\gamma$ of $\mathbb{F}^{\times}_q$, there exists a constant $C$ such that if $s\ge C$, $\frac{k-1}{s}\ge a$, and $n$ is sufficiently large, then any rate 
$R=\frac{k}{sn}$ folded Reed--Solomon code $\mathsf{FRS}^{(s,\gamma)}_{n,k}(\alpha_1,\alpha_2,\dots,\alpha_n)$ with appropriate evaluation points in $\mathbb{F}_q$ is $(\rho-\epsilon,\ell, L)$-list-recoverable, where
\[\rho=\frac{L+1-\ell}{L+1}\left(1-\frac{aR}{a-1}\right).\]
\end{conjecture}

\cref{conj:main} proposes a list-recovery analogue of the generalized Singleton bound $\rho\leq\tfrac{L}{L+1}(1-R)$ for list decoding ($\ell=1$) \cite{shangguan2020combinatorial}. For RS and folded RS codes, the list-decoding bound is achievable up to arbitrarily small slack \cite{brakensiek2023generic,gz23,agl24,CZ25}.

In the true list-recovery case where $\ell\ge 2$, we know very little about the correctness of this conjecture. Except for trivial special cases, prior to our work, the only known positive evidence for \cref{conj:main} was that \cite[Theorem 1.8]{CZ25} affirmed it in the special case $\ell=2,L=3$. No negative evidence against the correctness of \cref{conj:main} is known.

In this paper, we provide partial results toward the aforementioned conjecture, which we hope will serve as a step toward its full resolution.

All results in this paper also hold for univariate multiplicity codes. To simplify the presentation, we
choose to focus only on folded Reed--Solomon codes in this paper.

\subsection{Main Results}
Our first main result is a zero-error list-recovery bound for folded Reed--Solomon codes.
\begin{theorem}\label{thm:eps-version}
Let $\eps>0$, and let $\ell,a\ge2$ and $s>\ell^a/\eps$ be integers. For a finite field $\F_q$ and a generator $\gamma$ of $\F_q^\times$, any appropriate $s$-folded RS code $\mathsf{FRS}^{(s,\gamma)}_{n,k}(\alpha_1,\alpha_2,\dots,\alpha_n)$ with rate $R\leq \frac{a-1}{a+\ell-2}-\eps$ is $(0,\ell, \ell^a-1)$-list-recoverable.
\end{theorem}
\cref{thm:eps-version} follows from its parameterized version, \cref{thm:list-recovery}, by taking $s>\ell^a/\epsilon$.

The negative result in \cite{CZ25} implies that if the rate $R>\frac{a-1}{a}$, achieving $(0,\ell,\ell^a-1)$-list-recoverability is impossible for folded Reed--Solomon codes. When $\ell=2$, \cref{thm:eps-version} shows that folded Reed--Solomon codes with rate arbitrarily close to $\frac{a-1}{a}$ are $(0,\ell,\ell^a-1)$-list-recoverable, which confirms \cref{conj:main} in this special case $\rho=0,\ell=2$.

\cref{thm:eps-version} gives a new positive list-recovery bound for folded RS codes when the radius is $\rho=0$, a setting known as zero-error list recovery. We comment that zero-error list recovery has significant applications in pseudorandomness. A key connection is that zero-error list-recoverable (linear) codes are equivalent to (linear) lossless condensers with corresponding parameters \cite{guruswami2009Unbalanced}.

Our second main result shows that, at some rate arbitrarily close to any prescribed $R\in(0,\tfrac{a-1}{a})$, achieving the exact bound in \cref{conj:main}, without slack, requires exponential alphabet size.
\begin{theorem}[Simplified Version of \cref{thm:general-alphabet-size-bound}]\label{thm:simplify}
Fix integers $s\geq1$, $L\geq\ell\geq2$, and set $a:=\lceil\log_\ell(L+1)\rceil$. Let $R\in\left(0,\frac{a-1}{a}\right)$ and $\delta\in(0,1)$. For all sufficiently large $n$, let $\F_q$ be a finite field with $q>sn$, $\gamma$ a generator of $\F_q^\times$, and $\alpha_1,\ldots,\alpha_n\in\F_q$ appropriate evaluation points. There exists an integer $k'\in((1-\delta)Rsn,Rsn]$ such that, writing $R'=k'/(sn)$, if $\mathsf{FRS}^{(s,\gamma)}_{n,k'}(\alpha_1,\ldots,\alpha_n)$ is $\left(\frac{L+1-\ell}{L+1}\left(1-\frac{a}{a-1}R'\right),\ell,L\right)$-list-recoverable, then
\[
q^s\geq 2^{\Omega_{\ell,L,R,\delta}(n)}.
\]
\end{theorem}

We give the first nontrivial lower bound on the alphabet size toward \cref{conj:main}. In the list-decoding setting (which corresponds to the special case $\ell=1$ of list-recovery), related alphabet-size lower bounds were proved in \cite{bdg23, alrabiah2024ag}. Specifically, \cite{alrabiah2024ag} proves that any code achieving the exact generalized Singleton bound $\left(\tfrac{L}{L+1}(1-R), L\right)$ must have alphabet size $\exp(\Omega(n))$.

Thus \cref{thm:simplify} gives an analogous alphabet-size obstruction for folded RS list recovery at the radius proposed in \cref{conj:main}.

\begin{remark}
Recent works \cite{CZ25, li2025near} demonstrate a sharp separation between linear and nonlinear codes in the list-recovery setting: while a random (nonlinear) code achieves $\eps$-slacked list-recovery capacity with list size $O(\ell/\epsilon)$ with high probability, any linear code requires list size at least $\ell^{\Omega(R/\epsilon)}$. The alphabet size lower bound of \cite{alrabiah2024ag} applies to \emph{all} codes, whereas the radius bound in \cite{CZ25} is specific to linear codes. This separation suggests that the techniques of \cite{alrabiah2024ag} are unlikely to extend to the radius bound proposed in Conjecture \ref{conj:main}, which is formulated for folded RS codes and, more generally, linear codes.
\end{remark}

\paragraph{Follow-up Work} In two follow-up works \cite{BCDZ25lr,BCDZ25subspace}, building on ideas from \cref{thm:eps-version}, the authors affirm \cref{conj:main} for $\rho=0$ and all $\ell\ge2$. This significantly improves \cref{thm:eps-version}, which only affirms \cref{conj:main} when $\rho=0,\ell=2$. Moreover, \cite{BCDZ25lr} shows that explicit folded RS codes are $(1-R-\epsilon,\ell,(\tfrac{\ell}{R+\epsilon})^{O(1+R/\epsilon)})$-list-recoverable. The general case of \cref{conj:main} when $\rho>0$ is still open.

\section{Preliminaries}

The zero-error list-recovery argument uses the strong subspace design bound in \cref{lem:subspace-design}. We derive it from the folded Wronskian, which relates linear independence to polynomial vanishing. Throughout this section, fix an appropriate $s$-folded RS code of message dimension $k$, with encoder $\mathcal C$ and folding generator $\gamma$. All spans and dimensions are over $\F_q$.

\begin{definition}[Folded Wronskian, see \cite{guruswami2016explicit}]
For $d\geq1$ and polynomials $f_1,\ldots,f_d\in\F_q[X]$, their $\gamma$-folded Wronskian is the $d\times d$ matrix
\[
W_\gamma(f_1,\ldots,f_d)(X):=
\begin{pmatrix}
f_1(X) & \cdots & f_d(X) \\
f_1(\gamma X) & \cdots & f_d(\gamma X) \\
\vdots & \ddots & \vdots \\
f_1(\gamma^{d-1}X) & \cdots & f_d(\gamma^{d-1}X)
\end{pmatrix}.
\]
\end{definition}

Here $d$ counts the polynomials, while $s$ remains the folding parameter. The determinant detects linear independence of low-degree polynomials.

\begin{lemma}[Folded Wronskian criterion, see \cite{guruswami2016explicit,forbes2012identity}]\label{lem:wrons_ind}
Let $k<q$, and let $f_1,\ldots,f_d\in\F_q[X]_{\leq k-1}$. Then $f_1,\ldots,f_d$ are linearly independent over $\F_q$ if and only if $\det W_\gamma(f_1,\ldots,f_d)(X)\neq0$.
\end{lemma}

Changing the basis of a polynomial subspace changes its Wronskian determinant only by a nonzero scalar.

\begin{claim}\label{clm:wrons_equal}
Let $f_1,\ldots,f_d$ and $g_1,\ldots,g_d$ be two bases of the same subspace of $\F_q[X]$. Then, for some $c\in\F_q^\times$,
\[
\det W_\gamma(f_1,\ldots,f_d)(X)
=c\det W_\gamma(g_1,\ldots,g_d)(X).
\]
\end{claim}

\begin{proof}
If $\Span(f_1,\dots,f_d)=\Span(g_1,\dots,g_d)$, then $W_\gamma\left(f_1, \ldots, f_d\right)(X)$ and $W_\gamma\left(g_1, \ldots, g_d\right)(X)$ can be transformed into each other by column operations, and column operations preserve the determinant up to a nonzero scalar.
\end{proof}

Vanishing at a folded coordinate forces roots of the Wronskian determinant, with multiplicity determined by the dimension of the vanishing subspace.

\begin{claim}\label{clm:wrons_sub}
Let $f_1,\ldots,f_d\in\F_q[X]_{\leq k-1}$ be linearly independent, with $d\leq s$. Suppose $g_1,\ldots,g_r$ are linearly independent elements of $\Span(f_1,\ldots,f_d)$, where $1\leq r\leq d$, and $\mathcal C(g_j)[i]=0^s$ for every $j\in[r]$ at some coordinate $i\in[n]$. Then $\alpha_i,\gamma\alpha_i,\ldots,\gamma^{s-d}\alpha_i$ are roots of $\det W_\gamma(f_1,\ldots,f_d)(X)$, each with multiplicity at least $r$.
\end{claim}

\begin{proof}
Extend $g_1,\ldots,g_r$ to a basis $g_1,\ldots,g_d$ of $\Span(f_1,\ldots,f_d)$. By \cref{clm:wrons_equal}, it suffices to prove the claim for this basis. For $0\leq u\leq s-d$, $v\in[d]$, and $j\in[r]$, the $(v,j)$-entry $g_j(\gamma^{v-1}X)$ vanishes at $\gamma^u\alpha_i$, since $0\leq u+v-1\leq s-1$. Thus each of these points is a common root of every entry in the first $r$ columns. Expanding the determinant shows that each has multiplicity at least $r$.
\end{proof}

The preceding facts give the strong subspace design bound of \cite{guruswami2016explicit}.

\begin{corollary}[Strong subspace design bound]\label{lem:subspace-design}
For an appropriate $s$-folded RS code of message dimension $k$, let
\[
H_i:=\{f\in\F_q[X]_{\leq k-1}:\mathcal C(f)[i]=0^s\},\qquad i\in[n].
\]
For every $d$-dimensional subspace $V\subseteq\F_q[X]_{\leq k-1}$ with $1\leq d\leq s$, we have
\[
\sum_{i=1}^n\dim(V\cap H_i)\leq\frac{d(k-1)}{s-d+1}.
\]
\end{corollary}

\begin{proof}
Choose a basis $f_1,\ldots,f_d$ of $V$ and put $P(X)=\det W_\gamma(f_1,\ldots,f_d)(X)$. By \cref{lem:wrons_ind}, $P$ is nonzero and has degree at most $d(k-1)$. By \cref{clm:wrons_sub}, coordinate $i$ contributes at least $(s-d+1)\dim(V\cap H_i)$ roots of $P$, counted with multiplicity. The evaluation points are appropriate, so the roots from different coordinates are distinct. Summing their multiplicities proves the bound.
\end{proof}

\section{Near-Optimal Zero-Error List Recoverability of Explicit Folded RS Codes}

We now prove our zero-error list-recovery bound. The following theorem implies \cref{thm:eps-version} for $s>\ell^a/\eps$.

\begin{theorem}\label{thm:list-recovery}
Let $\ell,a\ge2$ and $s\ge L=\ell^a-1$ be integers, and suppose
\[
R<\min_{a\leq t\leq L}\left\{\frac{(s-t+1)\max\{t-\ell+1,a-1\}}{st}\right\}.
\]
Then any appropriate $s$-folded RS code $\mathsf{FRS}^{(s,\gamma)}_{n,k}(\alpha_1,\alpha_2,\dots,\alpha_n)$ with rate $\frac{k}{sn}=R$ is $(0,\ell,L)$-list-recoverable.
\end{theorem}

\begin{remark}
In the limit $s\to\infty$, the rate condition in \cref{thm:list-recovery} becomes
\[
R<\min_{a\leq t\leq L}\left\{\frac{\max\{t-\ell+1,a-1\}}{t}\right\}
=\frac{a-1}{a+\ell-2}=1-\frac{\ell-1}{a+\ell-2}.
\]
Both arguments use the strong subspace design property: ours bounds kernel dimensions directly, whereas \cite{BCDZ25lr} uses the discrete Brascamp--Lieb inequality to obtain $R<1-\tfrac{\log\ell}{\log(L+1)}=\tfrac{a-1}{a}$ in the same limit.
\end{remark}

We first describe the configurations that obstruct zero-error list recovery.

\begin{definition}[$\ell$-enveloped sets]\label{def:ell-enveloped}
Fix an appropriate $s$-folded RS code of length $n$ and dimension $k$, with encoder $\mathcal C$. A finite set $\mathcal F\subseteq\F_q[X]_{\leq k-1}$ is \emph{$\ell$-enveloped} if
\[
\bigl|\{\mathcal C(f)[i]:f\in\mathcal F\}\bigr|\leq\ell
\qquad\text{for every }i\in[n].
\]
\end{definition}
Thus an $\ell$-enveloped set of $L+1$ distinct message polynomials witnesses failure of $(0,\ell,L)$-list recoverability.

To prove the theorem, suppose there are $L+1=\ell^a$ such polynomials. Translate the set to contain $0$, and let $t\leq L$ be the dimension of its span $V$. At each coordinate $i$, the kernel of $f\mapsto\mathcal C(f)[i]$ on $V$ has dimension at least $t-\ell+1$, since its image is spanned by at most $\ell-1$ nonzero symbols. On the other hand, \cref{lem:subspace-design} gives an upper bound on the sum of the kernel dimensions.

The following lemma bounds the size of an $\ell$-enveloped set in terms of its affine dimension. We use it to bound $\dim V$ from below and, together with the pigeonhole principle, to obtain a further lower bound on the kernel dimensions. Comparing these bounds gives a contradiction under the stated condition on $R$.

\begin{lemma}\label{clm:impossible}
Let $\ell\geq2$ and $c\geq1$ be integers. For an appropriate $s$-folded RS code of message dimension $k$, let $f_1,\ldots,f_h\in\F_q[X]_{\leq k-1}$ be $h$ distinct polynomials, where $h>\ell^c$. If these polynomials lie in an affine subspace of dimension at most $c$, then they are not $\ell$-enveloped.
\end{lemma}

\begin{proof}
We prove the lemma by induction on $c$. Suppose for a contradiction that $f_1,\ldots,f_h$ are $\ell$-enveloped. For each $i\in[n]$, let $T_i:=\{\mathcal C(f_j)[i]:j\in[h]\}$, so $|T_i|\leq\ell$. We may assume $f_1=0$ by subtracting $f_1$ from all the polynomials and $\mathcal C(f_1)[i]$ from all the symbols in $T_i$. This preserves distinctness, affine dimension, and the $\ell$-enveloped property.

When $c=1$, the polynomials are collinear. Write $f_j=a_jg$ for $j\in[h]$, where $a_1,\ldots,a_h\in\F_q$ are distinct and $g$ is a nonzero polynomial of degree at most $k-1$. Since $\deg g<sn$ and the evaluation points are appropriate, there is some $\beta=\gamma^u\alpha_i$, with $i\in[n]$ and $0\leq u<s$, such that $g(\beta)\neq0$. Thus $f_1(\beta),\ldots,f_h(\beta)$ are distinct, and consequently $|T_i|=h>\ell$, a contradiction.

Now let $c>1$, and put $V=\mathrm{Span}(f_1,\ldots,f_h)$, so $\dim V\leq c$. Since $f_1\neq f_2$ and $\deg(f_2-f_1)<sn$, there exist $i\in[n]$ and $u\in\{0,\ldots,s-1\}$ such that $f_1(\gamma^u\alpha_i)\neq f_2(\gamma^u\alpha_i)$. Hence $\mathcal C(f_1)[i]\neq\mathcal C(f_2)[i]$. Let
\[
W:=\{f\in V:\mathcal C(f)[i]=0^s\}
\]
be the subspace of polynomials vanishing at this coordinate. Since $f_1=0$, we have $W\subsetneq V$ and hence $\dim W\leq c-1$. Choose a complementary subspace $W'\subseteq V$, so $W+W'=V$ and $W\cap W'=\{0\}$. Each $f_j$ has a unique expression
\[
f_j=f'_j+g_j,\qquad f'_j\in W,\quad g_j\in W'.
\]
Since $\mathcal C(f'_j)[i]=0^s$, we have $\mathcal C(f_j)[i]=\mathcal C(g_j)[i]$. Moreover, if $g_u\neq g_v$, then $g_u-g_v$ is a nonzero element of $W'$ and therefore does not lie in $W$. It follows that $\mathcal C(g_u)[i]\neq\mathcal C(g_v)[i]$. Thus
\[
|\{g_j:j\in[h]\}|=|\{\mathcal C(f_j)[i]:j\in[h]\}|=|T_i|\leq\ell.
\]

By the pigeonhole principle, there is some $g\in W'$ for which more than $\ell^{c-1}$ of the polynomials satisfy $g_j=g$. These polynomials are distinct and $\ell$-enveloped, and all lie in the affine subspace $g+W$, whose dimension is at most $c-1$. This contradicts the induction hypothesis.
\end{proof}

\begin{proof}[Proof of \cref{thm:list-recovery}]
Suppose the code is not $(0,\ell,L)$-list-recoverable. Then there are $L+1$ distinct message polynomials $f_0,\ldots,f_L$ that are $\ell$-enveloped. Translating by $-f_0$ preserves this property, so we may assume $f_0=0$. For each $i\in[n]$, set $T_i:=\{\mathcal C(f_j)[i]:0\leq j\leq L\}$; then $|T_i|\leq\ell$.

Let $V=\mathrm{Span}(f_1,\ldots,f_L)$ and $t=\dim V$. By \cref{clm:impossible}, $a\leq t\leq L\leq s$. For each $i\in[n]$, put $W_i:=\{f\in V:\mathcal C(f)[i]=0^s\}$. The bound in \cref{lem:subspace-design} gives
\[
\sum_{i=1}^n\dim W_i\leq\frac{t(k-1)}{s-t+1}.
\]

We claim that $\dim W_i\geq t-\ell+1$ for every $i\in[n]$. Suppose not, so $\dim W_i\leq t-\ell$. Since $f_1,\ldots,f_L$ span $V$, we can successively choose distinct indices $j_1,\ldots,j_\ell\in[L]$ such that
\[
f_{j_r}\notin W_i+\mathrm{Span}(f_{j_1},\ldots,f_{j_{r-1}}),\qquad r\in[\ell].
\]
Here the span is $\{0\}$ when $r=1$. Each choice is possible because the subspace on the right has dimension at most $\dim W_i+r-1<t$, and hence cannot contain all of $f_1,\ldots,f_L$. Put $j_0=0$. For any $0\leq u<v\leq\ell$, the choice of $f_{j_v}$ implies $f_{j_v}-f_{j_u}\notin W_i$, so $\mathcal C(f_{j_v})[i]\neq\mathcal C(f_{j_u})[i]$. Therefore,
\[
|T_i|\geq|\{\mathcal C(f_{j_r})[i]:0\leq r\leq\ell\}|=\ell+1,
\]
contradicting $|T_i|\leq\ell$.

Another observation is that $\dim(W_i)\ge a-1$. We prove this claim in two cases: $W_i=V$ or $W_i\subsetneq V$. If $W_i=V$, then $\dim(W_i)=\dim(V)=t\ge a$. If $W_i\subsetneq V$, then we consider a complementary subspace $W'_i\subseteq V$ such that $W'_i+W_i=V$ and $W'_i\cap W_i=\{0\}$. For any $0\leq j\leq L$, there is a unique way to write $f_j$ as $f_j=h_j+g_j$, where $h_j\in W_i$ and $g_j\in W'_i$. Since $f_0,\dots,f_L$ are $\ell$-enveloped, for the same reason as in the proof of \cref{clm:impossible}, we have $|\{g_j\colon j\in\{0\}\cup[L]\}|\leq \ell$. Therefore, by the pigeonhole principle there must be some $g\in W'_i$ such that
\begin{compactitem}
\item $|\{f_j\colon j\in\{0\}\cup[L],g_j=g\}|\ge \ell^{a-1}>\ell^{a-2}$.
\item $\{f_j\colon j\in\{0\}\cup[L],g_j=g\}$ is a set of distinct $\ell$-enveloped polynomials.
\end{compactitem}

Since $\{f_j\colon j\in\{0\}\cup[L],g_j=g\}$ contains at least $\ell^{a-1}\ge2$ distinct polynomials in the affine subspace $g+W_i$, we have $\dim(W_i)\ge1$. Applying \cref{clm:impossible} with $c=\dim(W_i)$ implies $\dim(W_i)\ge a-1$.

Combining these lower bounds with \cref{lem:subspace-design}, we obtain
\[
n\max\{t-\ell+1,a-1\}
\leq\sum_{i=1}^n\dim W_i
\leq\frac{t(k-1)}{s-t+1}
<\frac{tRsn}{s-t+1},
\]
contrary to the assumed bound on $R$.
\end{proof}

\begin{remark}\label{rem:multiplicity-zero-error}
The proof uses only linearity, injectivity, and the bound in \cref{lem:subspace-design}. This bound also holds for $s$-order univariate multiplicity codes of message dimension $k$ over finite fields $\F_q$ with $\operatorname{char}(\F_q)>k$ and distinct evaluation points \cite{guruswami2016explicit,CZ25}. Hence \cref{thm:list-recovery,thm:eps-version} extend to these codes with the same rate conditions.
\end{remark}

\section{A Lower Bound on Alphabet Size for List-Recoverable Folded RS Codes}

In this section, we derive a nontrivial lower bound on the alphabet size for list-recoverable folded RS codes achieving the radius upper bound proposed in \cite{CZ25}. The theorem below states the result formally.

\begin{theorem}\label{thm:general-alphabet-size-bound}
Fix integers $s\geq1$, $L\geq\ell\geq2$, and set $a:=\lceil\log_\ell(L+1)\rceil$. Let $R\in\left(0,\frac{a-1}{a}\right)$ and $\delta\in(0,1)$. There exist positive constants $n_0=n_0(\ell,L,R,\delta)$, $\varepsilon_0=\varepsilon_0(\ell,L,R)$, $C_1=C_1(\ell,L,R,\delta)$ and $C_2$ with the following property.

For any integer $n\geq n_0$, let $\F_q$ be a finite field with $q>sn$, $\gamma$ a generator of $\F_q^\times$, and $\alpha_1,\ldots,\alpha_n\in\F_q$ appropriate evaluation points. For each $\varepsilon\in[0,\varepsilon_0]$, there exists an integer $k'\in((1-\delta)Rsn,Rsn]$ with the following property. Let $R':=k'/(sn)$. If $\mathsf{FRS}^{(s,\gamma)}_{n,k'}(\alpha_1,\ldots,\alpha_n)$ is $\left(\frac{L+1-\ell}{L+1}\left(1-\frac{a}{a-1}R'\right)-\varepsilon,\ell,L\right)$-list-recoverable, then
\[
q^s\geq 2^{\frac{C_1 n}{\varepsilon n+C_2}}.
\]
\end{theorem}

\cref{thm:general-alphabet-size-bound} shows that a large alphabet size is sometimes necessary for list-recoverable folded RS codes with $\rho=\frac{L+1-\ell}{L+1}\left(1-\frac{a}{a-1}R'\right)$ and list size $L$. In particular, setting $\varepsilon=0$ gives the lower bound $|\Sigma|\geq 2^{\Omega_{\ell,L,R,\delta}\left(n\right)}$ as stated in \cref{thm:simplify}.

Our approach builds on \cite{alrabiah2024ag,CZ25}. In \cite{alrabiah2024ag}, one finds many codewords agreeing with a fixed codeword on large coordinate sets. When the alphabet is small, a pigeonhole argument forces two of them to coincide on an additional set $I_0$, providing extra agreement.

This approach does not directly give the larger agreement required at the radius in \cite{CZ25}: two distinct folded RS codewords can agree on at most $(k'-1)/s$ coordinates. We therefore use the construction of \cite{CZ25}, where polynomials associated with disjoint sets $T_1,\ldots,T_a$ generate $\ell^a$ codewords taking at most $\ell$ symbols at every coordinate of $T=\bigcup_{j=1}^a T_j$.

The main difficulty is to make two of these codewords coincide on $I_0$ while keeping the sets $T_1,\ldots,T_a$ disjoint and sufficiently large. A direct pigeonhole argument may give overlapping sets; removing their intersection preserves the collision but may leave too few coordinates. We resolve this by iterating on unused coordinates and merging the disjoint pieces until they reach the required size.

We then choose $L+1$ codewords including the colliding pair. On $T$, include all their symbols in the received table. Elsewhere, distribute agreement cyclically among the $L+1$ codewords, treating the colliding pair as one group on $I_0$. This raises each codeword's share on $I_0$ from $\ell/(L+1)$ to $\ell/L$, giving the required violation of list recoverability.

We use the following elementary observation to distribute agreement among the codewords.

\begin{lemma}\label{lem:even-distribution}
Let $\Lambda,\lambda,\mu$ be positive integers with $\lambda\leq\Lambda$. For each $i\in[\mu]$, define
\[
M_i:=\left\{((i-1)\lambda+j-1)\bmod\Lambda+1:j\in[\lambda]\right\}.
\]
Then $|M_i|=\lambda$, and every element of $[\Lambda]$ belongs to either $\lfloor\lambda\mu/\Lambda\rfloor$ or $\lceil\lambda\mu/\Lambda\rceil$ of the sets $M_1,\ldots,M_\mu$.
\end{lemma}

\begin{proof}
The sets $M_i$ are consecutive blocks of length $\lambda$ in the cyclic sequence $1,\ldots,\Lambda,1,\ldots,\Lambda,\ldots$, with no repetition within a block since $\lambda\leq\Lambda$. Among the first $\lambda\mu$ entries, each element occurs either $\lfloor\lambda\mu/\Lambda\rfloor$ or $\lceil\lambda\mu/\Lambda\rceil$ times, proving the claim.
\end{proof}

For $A\subseteq[n]$, set
\[
f_A(X):=\prod_{i\in A}\prod_{j=0}^{s-1}\bigl(X-\gamma^j\alpha_i\bigr),
\]
with $f_{\varnothing}=1$.

\begin{proof}[Proof of \cref{thm:general-alphabet-size-bound}]
Let
\[
K:=\frac{Rn}{a-1},\qquad
\eta:=1-\frac{aR}{a-1}>0,\qquad
h:=\left\lfloor\frac{\delta K}{2}\right\rfloor.
\]
Let $m$ be a positive integer with $m\leq\eta n/2$, to be chosen below, and fix $I_0=[m]$. Put
\[
N:=n-m-2\lceil(1-\delta)K\rceil.
\]
We show that if $q^{sm}<\binom{N}{h}$, then there is an admissible $k'$ for which the claimed list-recovery property fails.

Start with $T_1=T_2=\varnothing$. As long as $|T_1|=|T_2|<(1-\delta)K$, consider all $h$-element subsets of
\[
[n]\setminus(I_0\cup T_1\cup T_2).
\]
There are at least $\binom{N}{h}$ such subsets, while their associated codewords have at most $q^{sm}$ possible restrictions to the coordinates in $I_0$, denoted by
\[
\left.\mathcal C(f_A)\right|_{I_0}\in(\F_q^s)^m.
\]
By the pigeonhole principle, there are two distinct subsets $A_1,A_2$ such that
\[
\left.\mathcal C(f_{A_1})\right|_{I_0}
=\left.\mathcal C(f_{A_2})\right|_{I_0}.
\]
Set $B_1=A_1\setminus A_2$ and $B_2=A_2\setminus A_1$. The factor $f_{A_1\cap A_2}$ is nonzero at every folded evaluation point over $I_0$, since the evaluation points are appropriate and $A_1\cup A_2$ is disjoint from $I_0$. Cancelling this factor gives $\left.\mathcal C(f_{B_1})\right|_{I_0}=\left.\mathcal C(f_{B_2})\right|_{I_0}$.

Replace $T_j$ by $T_j\cup B_j$ for $j=1,2$. The sets remain disjoint, and the equality $\left.\mathcal C(f_{T_1})\right|_{I_0}=\left.\mathcal C(f_{T_2})\right|_{I_0}$ is preserved, since $f_{T_j\cup B_j}=f_{T_j}f_{B_j}$. Each iteration increases their common size by an integer between $1$ and $h$. Thus the process terminates with
\[
(1-\delta)K\leq d:=|T_1|=|T_2|<(1-\delta)K+h
\leq(1-\delta/2)K.
\]
Set $k':=s(a-1)d+1$. For sufficiently large $n$, we have
\[
(1-\delta)Rsn<k'\leq Rsn.
\]
Moreover, $ad+m<aK+\eta n/2<n$, so we can choose pairwise disjoint sets $T_3,\ldots,T_a$ of size $d$ outside $I_0\cup T_1\cup T_2$.

Let $T=\bigcup_{j=1}^a T_j$ and define
\[
g_j(X):=f_{T\setminus T_j}(X),\qquad j\in[a].
\]
Each $g_j$ has degree $s(a-1)d=k'-1$. The polynomials $g_1,\ldots,g_a$ are linearly independent over $\F_q$: evaluating a relation $\sum_{v=1}^a\lambda_vg_v=0$ at any $\alpha_i$ with $i\in T_j$ gives $\lambda_jg_j(\alpha_i)=0$, and hence $\lambda_j=0$. Since $q\geq sn\geq n>\ell$, choose $\ell$ distinct elements of $\F_q$, including $0$ and $1$, and let $\mathcal B$ be their set. Define
\[
G:=\left\{\sum_{j=1}^a\beta_jg_j:\beta_1,\ldots,\beta_a\in\mathcal B\right\}.
\]
By linear independence, $|G|=\ell^a\geq L+1$; moreover, $g_1,\ldots,g_a\in G$ because $0,1\in\mathcal B$. Also,
\[
g_1-g_2=f_{T_3\cup\cdots\cup T_a}(f_{T_2}-f_{T_1}).
\]
Hence $\mathcal C(g_1)$ and $\mathcal C(g_2)$ agree on $I_0$. Select $L+1$ distinct polynomials $h_1,\ldots,h_{L+1}\in G$ with $h_L=g_1$ and $h_{L+1}=g_2$, and write $c_j=\mathcal C(h_j)$. These codewords are distinct because the encoding is injective on polynomials of degree less than $k'\leq Rsn<sn$.

Partition the coordinates into $T$, $U:=[n]\setminus(T\cup I_0)$, and $I_0$. We construct $S=S_1\times\cdots\times S_n$ on these three sets as follows.
\begin{itemize}
    \item On $T$, include all symbols appearing among $c_1,\ldots,c_{L+1}$. That is, for each $i\in T$, set
    \[
    S_i:=\{c_j[i]:j\in[L+1]\}
    =\{\mathcal C(h_j)[i]:j\in[L+1]\}.
    \]
    \item On $U$, write $U=\{u_1,\ldots,u_{|U|}\}$ with $u_1<\cdots<u_{|U|}$. For each $v\in[|U|]$, set
    \[
    M_v:=\left\{((v-1)\ell+j-1)\bmod(L+1)+1:j\in[\ell]\right\},
    \qquad S_{u_v}:=\{c_j[u_v]:j\in M_v\}.
    \]
    \item On $I_0=[m]$, treat $c_L$ and $c_{L+1}$ as one group, represented by $c_L$. For each $i\in[m]$, set
    \[
    M'_i:=\left\{((i-1)\ell+j-1)\bmod L+1:j\in[\ell]\right\},
    \qquad S_i:=\{c_j[i]:j\in M'_i\}.
    \]
    When $L\in M'_i$, both $c_L[i]$ and $c_{L+1}[i]$ belong to $S_i$, since these two symbols are equal.
\end{itemize}

We first check that $|S_i|\leq\ell$ for every $i\in[n]$. If $i\in T_v$, then $\mathcal C(g_j)[i]=0^s$ for $j\neq v$. Since $h_1,\ldots,h_{L+1}\in G$, it follows that
\[
S_i\subseteq\{\beta\mathcal C(g_v)[i]:\beta\in\mathcal B\},
\]
so $|S_i|\leq|\mathcal B|=\ell$. On $U$ and $I_0$, each defining index set $M_v$ or $M'_i$ has exactly $\ell$ elements, since $\ell\leq L<L+1$, and hence $|S_i|\leq\ell$ as well.

We now apply \cref{lem:even-distribution} to $M_1,\ldots,M_{|U|}$ with $(\Lambda,\lambda,\mu)=(L+1,\ell,|U|)$. Every $j\in[L+1]$ belongs to at least $\lfloor\ell|U|/(L+1)\rfloor$ of these sets, so $c_j$ agrees with $S$ on at least this many coordinates of $U$. Applying the lemma to $M'_1,\ldots,M'_m$ with $(\Lambda,\lambda,\mu)=(L,\ell,m)$ gives at least $\lfloor\ell m/L\rfloor$ agreements on $I_0$ for each of $c_1,\ldots,c_L$. The same holds for $c_{L+1}$ because it agrees with $c_L$ on $I_0$. Finally, all $L+1$ codewords agree with $S$ throughout $T$ by construction. Writing $r=n-|T|$, so that $|U|=r-m$, we obtain, for every $j\in[L+1]$,
\[
\mathrm{Agr}(c_j,S)
\geq |T|+\left\lfloor\frac{\ell(r-m)}{L+1}\right\rfloor
+\left\lfloor\frac{\ell m}{L}\right\rfloor.
\]
Since $|T|=a(k'-1)/(s(a-1))$, the preceding inequality gives
\begin{align*}
\mathrm{dist}(c_j,S)
&\leq\frac{L+1-\ell}{L+1}r-\frac{\ell m}{L(L+1)}+2\\
&=\frac{L+1-\ell}{L+1}\left(n-\frac{ak'}{s(a-1)}\right)
+\frac{a(L+1-\ell)}{s(a-1)(L+1)}-\frac{\ell m}{L(L+1)}+2.
\end{align*}
We therefore choose
\[
m:=\left\lceil\frac{L(L+1)}{\ell}
\left(\varepsilon n+2+\frac{a(L+1-\ell)}{s(a-1)(L+1)}\right)\right\rceil.
\]
It follows that every $j\in[L+1]$ satisfies
\[
\mathrm{dist}(c_j,S)
\leq\frac{L+1-\ell}{L+1}\left(n-\frac{ak'}{s(a-1)}\right)-\varepsilon n.
\]
Thus these $L+1$ words violate the required list-recovery bound. The construction also applies when $L=\ell$: all $L$ groups are then selected at every coordinate of $I_0$.

It remains to check the coordinate budget and count the subsets. Let $b=L(L+1)/\ell$. Since $a/(s(a-1))\leq2$, our choice of $m$ satisfies
\[
m\leq b(\varepsilon n+5).
\]
We then choose $\varepsilon_0=\eta/(4b)$. For sufficiently large $n$ and $0\leq\varepsilon\leq\varepsilon_0$, we have $m\leq\eta n/2$, as required. Also, $h=\Theta_{\ell,L,R,\delta}(n)$. Since $a\geq2$ and $n-aK=\eta n$, we have
\begin{align*}
N-h&\geq n-m-2K+\frac{3\delta K}{2}-2\\
&=\eta n-m+\left(a-2+\frac{3\delta}{2}\right)K-2\\
&\geq\frac{\eta n}{2}-2\geq\frac{\eta n}{4},
\end{align*}
where the last inequality holds for $n\geq8/\eta$.
Consequently, for some $c=c(\ell,L,R,\delta)>0$,
\[
\binom{N}{h}\geq\left(\frac{N}{h}\right)^h\geq 2^{cn}.
\]
If $q^{sm}<\binom{N}{h}$, the construction above supplies an admissible $k'$ for which list recovery fails. Otherwise, the desired alphabet bound holds independently of the choice of $k'$:
\[
|\F_q^s|=q^s\geq 2^{cn/m}
\geq 2^{\frac{(c/b)n}{\varepsilon n+5}}.
\]
Taking $C_1=c/b$ and $C_2=5$ completes the proof.
\end{proof}

\begin{remark}
The conclusion of \cref{thm:general-alphabet-size-bound} also holds for $s$-order univariate multiplicity codes over any finite field $\F_q$ with $q>sn$, alphabet $\F_q^s$, and distinct evaluation points $\alpha_1,\ldots,\alpha_n\in\F_q$. It suffices to redefine $f_A(X):=\prod_{i\in A}(X-\alpha_i)^s$; with Hasse derivatives, the same construction works in every characteristic.
\end{remark}

\section*{Acknowledgment}
The authors used GPT-6 Astra to assist in simplifying the proof of
Theorem~\ref{thm:general-alphabet-size-bound} and to improve wording throughout the paper.
The main results and original proofs were obtained by the authors.
The authors verified all AI-assisted revisions and take full
responsibility for the content of the paper.

\bibliography{main}

\end{document}